**Titre principal (Main Title) :**
*Modélisation endo-irréversible des moteurs thermomécaniques avec nouveau concept d'action de production d'entropie*
Endo-irreversible thermo-mechanical engine with new concept of entropy production action coefficient

**Auteurs (Authors) :** Michel FEIDT, Renaud FEIDT

Résumé (Summary) : La Thermostatique des moteurs de CARNOT a été prolongée par des travaux plus récents avec modèle endoréversible. Notre modèle suppose l'exo-réversibilité du convertisseur au contact des sources et puits pour déterminer de nouvelles limites à la conversion thermomécanique.
Nous proposons une expression fonctionnelle de production d'entropie liée aux durées des transformations du cycle. Cette approche analyse les conséquences en énergie, entropie, puissance. On introduit un nouveau concept d'action de production d'entropie dont on déduit trois optimums : le maximum d'énergie rapporté aux durées de transformation et celui associé à l'équipartition des actions entropiques, l'optimisation de la puissance pour une période de cycle.

*Thermostatics of CARNOT engines has been extended by more recent research based on endo-reversible model. Our model assumes exo-reversibility but endo-irreversibility to determine new upper-bound to thermomechanical conversion.*
*We propose a functional expression of entropy production related to transformation cycle durations. This approach analyses the energy, entropy and power consequences. We introduce a new concept of entropy production actions that results in three optimums : maximum energy related to transformation durations, maximum energy associated with equipartition of entropy actions, optimal power for given period cycle.*

**Mots clés (Keywords) :** moteur thermomécanique, rendement, optimisation, énergie, puissance, Carnot
*(thermodynamics, efficiency, optimization, energy, power, Carnot engine)*

1. **Introduction**

La révolution industrielle a pris son essor conjointement à celui de la Thermodynamique. Si les préoccupations de l'ingénieur ont été importantes, il n'en reste pas moins que des recherches plus fondamentales ont accompagné le développement des moteurs thermiques (ou plus précisément thermomécaniques).

Sous l'influence probable de son père L. Carnot qui été mécanicien [1], S. Carnot a été le premier à faire le lien entre la mécanique et la thermique [2] dans un apport déterminant qui fait de lui un des pères fondateurs de la Thermodynamique de l'équilibre. On lui doit ainsi plusieurs apports décisifs [3] :
- La notion de cycle thermodynamique, dont le cycle qui porte son nom,
- L'extension de la notion de rendement mécanique, au rendement d'un moteur thermomécanique (on dit maintenant rendement au sens du premier principe de la thermodynamique, $\eta_I$), forme adimensionnelle d'efficacité.

S. Carnot a montré que ce rendement est borné supérieurement par ce qui est appelé, le rendement de Carnot, $\eta_C$. Ce rendement caractérise, les moteurs dithermes, fonctionnant de façon quasi statique (Thermodynamique de l'équilibre, T.E.). En supposant la source thermique isotherme à T$_{HS}$ (thermostat), de même que le puits thermique isotherme à T$_{CS}$ < T$_{HS}$, il obtient :

$$\eta_C = 1 - \frac{T_{CS}}{T_{HS}} \qquad (1)$$

Depuis lors de nombreux travaux concernent le cycle de Carnot, ou les moteurs de Carnot (on recense ainsi à ce jour 268 références avec ces mots clés sur les 3 dernières années). Le cadre de ces travaux déborde largement l'ingénierie et la physique classique, pour aller vers des moteurs quantique, moléculaire, photonique… Des modèles autres quittent les petites échelles (nano physique) pour considérer de grosses structures (en particulier les trous noirs), ou le vivant.

Pour ce qui concerne la présente proposition, elle restera centrée sur l'approche phénoménologique de la thermodynamique classique aux échelles mésoscopiques (humaines).

Dans ce cadre, il faut noter un renouveau des travaux visant à considérer l'influence du hors équilibre des transformations de cycles. L'article le plus souvent cité comme marquant ce renouveau est l'article de Curzon-Ahlborn [4] paru en 1975. En fait, comme nous l'avons montré [5], ce travail avait été précédé par d'autres travaux en 1957 [6,7] et même avant [8].

La majorité de ces travaux suppose des écarts de température entre la source et le moteur (éventuellement entre le moteur et le puits), mais au final, considère un moteur endoréversible (sans irréversibilité interne) pour l'optimisation.

De plus l'optimisation porte principalement sur la puissance $\dot{W}$ (en watt) du moteur. Le choix de cette fonction objectif est logique, vu que c'est l'effet utile d'un moteur. Néanmoins quelques précisions seront ajoutées dans cet article. Le résultat phare de l'optimisation est la mise en évidence du rendement du moteur endoréversible au maximum de puissance :

$$\eta_I(MAX\ \dot{W}) = 1 - \sqrt{\frac{T_{CS}}{T_{HS}}} \qquad (2)$$

Ce rendement est bien inférieur au rendement de CARNOT, mais le moteur fournit une certaine puissance (ce qui n'est pas le cas pour le modèle de Thermodynamique de l'équilibre). Depuis lors, des recherches ont montré que la relation (2) n'était pas générique [9,10,11]. Le présent article propose essentiellement deux choses :
- Contrairement à la majorité des propositions qui se focalisent sur les irréversibilités externes aux transferts de chaleur, nous développons le cas où les transferts de chaleur sont parfaits (exo-réversibilité), mais où les irréversibilités internes au convertisseur sont prises en compte quelles qu'elles soient. Il n'est pas fait appel à des modèles à faible dissipation [11] ou aux pseudo-cycles de Carnot [12]. Alors apparaît l'importance des entropies de transfert de chaleur (voir section 2)

- Nous considérerons 2 cas majeurs d'optimisation : soit en énergie W, soit en puissance moyenne $\overline{W}$ (et non pas puissance instantanée, comme nous justifierons (voir section 3))

Dans ces cas, l'influence de la forme des irréversibilités internes, sera particulièrement étudiée, ainsi que les conséquences qui en découlent.

La section 4 sera dédiée à la discussion, puis à la synthèse des résultats obtenus. Les perspectives qui en découlent seront proposées.

## 2. Modèle endo-irréversible de moteur de Carnot

Contrairement à la grande majorité des articles parus sur le sujet, nous présentons un modèle dont les irréversibilités sont internes au convertisseur et représentées de façon globale, comme il sera vu ci-après.

### 2.1 Modèles avec pertes thermiques

Toutefois, pour être complet, la figure 1 représente le moteur de CARNOT placé entre le thermostat chaud à $T_{HS}$ et le thermostat froid à $T_{CS}$. On remarque que ce schéma intègre la présence des pertes thermiques (à travers la structure matérielle) entre le point le plus chaud et le point le plus froid. La conductance de pertes thermiques équivalente au système $K_{LS}$ conduit classiquement, en supposant des transferts de chaleur linéaires, à l'expression de l'énergie thermique perdue $Q_{LS}$ :

$$Q_{LS} = K_{LS} \times \zeta \times (T_{HS} - T_{CS}) \qquad (3)$$

$\zeta$, durée de fonctionnement à préciser
$K_{LS}$, conductance des pertes thermiques en moyenne sur la durée de fonctionnement (W /K)

On écrira de façon générale :
$$Q_{LS} = G_{LS} \times (T_{HS} - T_{CS}) = T_{HS} \times \Delta S_{LS} \qquad (4)$$

$G_{LS}$, conductance énergétique de pertes thermiques (J.K$^{-1}$)
$\Delta S_{LS}$, entropie de transfert de pertes thermiques, à la source pendant $\zeta$

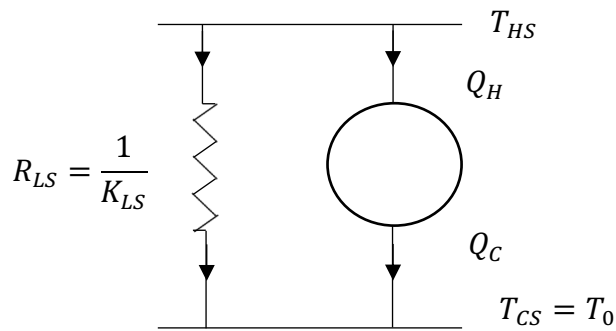

Figure 1 : représentation schématique générale du moteur de Carnot endo-irréversible avec pertes thermiques

Pour préserver le plus de généralité possible, nous conserverons la dépense énergétique $Q_{HS}$ sous la forme :

$$Q_{HS} = Q_H + Q_{LS} \qquad (5)$$

En posant $Q_H = T_{HS} \times \Delta S_H$, de même que $Q_C = T_{CS} \times \Delta S_C$, il vient :

$$Q_{HS} = T_{HS} \times (\Delta S_H + \Delta S_{LS}) \qquad (6)$$

Le bilan entropique au convertisseur s'écrit sur la même période :

$$\Delta S_C = \Delta S_H + \Delta S_I \qquad (7)$$

$\Delta S_I$, production interne d'entropie du convertisseur pendant $\zeta$

On en déduit, le travail mécanique produit sur la même durée, à l'aide du bilan d'énergie au convertisseur par différence de chaleur au bout chaud, puis froid du moteur ($Q_H - Q_C$) :

$$W = (T_{HS} - T_{CS}) \times \Delta S_H - T_{CS} \times \Delta S_I \qquad (8)$$

On retrouve ici que le maximum d'énergie mécanique est obtenu lorsque la production d'entropie s'annule (cycle totalement réversible). Mais on note que cette énergie dépend du potentiel thermique $\Delta T_S = T_{HS} - T_0$, et aussi de l'entropie de transfert de chaleur entre thermostats et convertisseur, que nous prendrons comme référence $\Delta S_H = \Delta S$.

L'expression du rendement au sens du premier principe s'exprime alors par :

$$\eta_I = \frac{W}{Q_{HS}} = \frac{(T_{HS} - T_{CS}) \times \Delta S - T_{CS} \times \Delta S_I}{T_{HS} \times (\Delta S + \Delta S_{LS})}$$

$$\eta_I = \frac{1}{1+d_{LS}} \times \left[1 - \frac{T_{CS}}{T_{HS}} \times (1 + d_I)\right] \qquad (9)$$

Ou

$$\eta_I = [\eta_C - (1 - \eta_C) \times d_{IS}] \times \frac{1}{1+d_{LS}}$$

Avec  $d_I$, degré d'irréversibilité du convertisseur $= \frac{\Delta S_I}{\Delta S}$

$d_{LS}$, degré de perte thermique (du système incluant source et puits) $= \frac{\Delta S_{LS}}{\Delta S}$

On voit sur l'expression (9) que les pertes thermiques introduisent une atténuation du rendement liée au pourcentage $d_{LS}$ (généralement de l'ordre de quelques pourcent). De même, les irréversibilités internes atténuent le rendement par l'intermédiaire de $d_I$, mais pondéré du facteur $\frac{(1-\eta_C)}{\eta_C}$. Les moteurs à faible écart de température sont de ce fait pénalisés.

On remarque d'ailleurs à degré d'irréversibilité fixé, l'existence d'un seuil de température de source conduisant à un rendement physiquement acceptable pour le moteur :

$$\eta_C > \frac{d_I}{1+d_I}$$

En général, $d_I$ est inférieur à 1.

Ou à températures données, l'existence d'un seuil d'irréversibilité :

$$d_I < \frac{\eta_C}{1-\eta_C}$$

L'étude liée au degré de perte thermique du système n'a de sens que pour un thermostat chaud d'énergie finie $Q_0$. Il est alors aisé de montrer que :

$$\Delta S = \frac{Q_0 - T_{HS} \times \Delta S_{LS}}{T_{HS}} \geq 0 \qquad (10)$$

Ce qui conduit, soit à une condition sur $\Delta S$

$$\Delta S \times (1 + d_{LS}) \leq \frac{Q_0}{T_{HS}} \qquad (a)$$

Soit à une condition sur $Q_0$

$$Q_0 \geq T_{HS} \times \Delta S_{LS} \qquad (b)$$

Mais la relation (10) montre aussi l'existence d'une température maximale atteignable en source d'énergie finie, ou <u>température de stagnation</u> $T_S$ (notion courante en énergétique solaire)

$$T_{HS} < \frac{Q_0}{d_{LS} \times \Delta S} = T_S \quad (11)$$

Cette relation peut aussi s'interpréter sous 2 autres formes :

$$\Delta S < \frac{Q_0}{T_{HS} \times d_{LS}} \qquad \text{ou} \qquad d_{LS} < \frac{Q_0}{T_{HS} \times \Delta S}$$

Elles induisent, à température $T_{HS}$ et énergie de source imposées, soit une limitation d'entropie de transfert vers le convertisseur, soit une limitation du degré des pertes thermiques.

Dans le paragraphe qui suit, nous supposerons par souci de simplicité, un moteur de CARNOT endo-irréversible sans pertes thermiques.

### 2.2 Modèle du moteur de Carnot endo-irréversible

Le cycle de CARNOT endo-irréversible est représenté sur la Figure 2. Il comporte classiquement quatre transformations : 2 isothermes et 2 adiabatiques irréversibles. Le cycle totalement réversible est représenté sur la Figure 2 comme le cycle 1 rev – 2 rev – 3 rev – 4 rev. Il y correspond en entrée du convertisseur $Q_{H\,rev}$ tel que :

$$Q_{H\,rev} = T_{HS} \times \Delta S_{H\,rev}$$

Pour ce même cycle réversible, mais au bout froid, il vient :

$$Q_{C\,rev} = T_{CS} \times \Delta S_{C\,rev} = T_{CS} \times \Delta S_{H\,rev}$$

L'énergie mécanique dans le cas réversible est alors exprimée sous la forme :

$$W_{rev} = Q_{H\,rev} - Q_{C\,rev} = (T_{HS} - T_{CS}) \times \Delta S_{H\,rev} = \Delta T_S \times \Delta S_{H\,rev} \quad (12)$$

Par soucis de simplification des notations, on posera ci-après (voir figure 2) :

$$\Delta S_{H\,rev} = \Delta S_{C\,rev} = \Delta S_{rev}$$

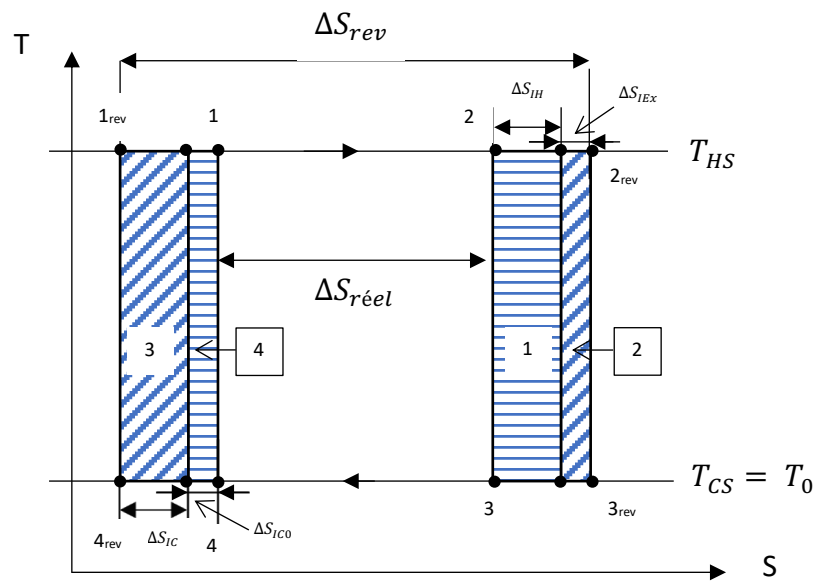

Figure 2 : représentation des cycles réversible et endo-irréversible de CARNOT, dans le diagramme (T,S)

Pour le cycle endo-irréversible, il apparaît à chaque transformation une production d'entropie : $\Delta S_{IH}$ pour l'isotherme haute température ; $\Delta S_{IEx}$ pour la détente adiabatique ; $\Delta S_{IC}$ pour l'isotherme basse température ; $\Delta S_{IC0}$ pour la compression adiabatique.
Chaque transformation dégrade de l'énergie mécanique = aire 1 pour l'isotherme haute température ; aire 2 pour la détente adiabatique, aire 3 pour l'isotherme basse température ; aire 4 pour la compression adiabatique. De ce fait, le travail mécanique du cycle endo-irréversible correspond à l'aire 1-2-3-4 soit :

$$W_{réel} = (T_{HS} - T_{CS}) \times \Delta S_{réel}$$

Or le bilan entropique du cycle (S, fonction d'état) fournit :

$$\Delta S_S = \Delta S_{rev} = \Delta S_{réel} + \Delta S_{IH} + \Delta S_{IC} + \Delta S_{IEx} + \Delta S_{IC0} = \Delta S_{réel} + \Delta S_I$$

D'où :

$$W_{réel} = (T_{HS} - T_{CS}) \times (\Delta S_S - \Delta S_I) \qquad (13)$$

$\Delta S_S$ , entropie de transfert de chaleur disponible à la source dans le cas réversible, ou entropie totale du système (source + convertisseur + puits)
$\Delta S_I$ , production totale d'entropie sur le cycle

$\Delta S_{réel}$ correspond en fait à l'entropie de conversion. On en déduit l'expression du rendement du moteur endo-irréversible par rapport à l'entropie de transfert du cas réversible :

$$\eta_I = \eta_C \times (1 - d_I) \quad (14)$$

Avec $d_I = \frac{\Delta S_I}{\Delta S_S}$ , le degré d'irréversibilité du convertisseur *(introduit par I. NOVIKOV)* [7]

La limite réversible ($d_I = 0$) restitue bien le rendement de Carnot de la Thermodynamique de l'équilibre, selon un schéma plus cohérent que celui proposé par WANG et HE [12] et conforme à des définitions antérieures [13].

### 2.3 Optimisation en temps fini de l'énergie du moteur

Nous supposerons ici que les 4 transformations du cycle se font en temps fini, selon le schéma voisin de celui utilisé dans certains travaux dont [12]. On suppose ainsi que chaque transformation s'effectue en une durée $\zeta_i$ , mais avec une durée totale finie $\zeta$ imposée pour le cycle telle que :

$$\sum_i \zeta_i = \zeta \qquad (15)$$

L'hypothèse la plus simple pour les dissipations consiste ,quelle que soit la transformation, à les supposer inversement proportionnelles à $\zeta_i$ :

$$\Delta S_{Ii} = \frac{C_{Ii}}{\zeta_i} \qquad (16)$$

En effet, cette hypothèse restitue bien la limite quasi-statique de la Thermodynamique de l'équilibre lorsque $\zeta_i \to \infty$ . On notera que les coefficients d'irréversibilité $C_{Ii}$ sont des quantités finies positives, ayant la dimension [J.s/K]. Nous les nommons **action de production d'entropie** (nouveau concept analogue à l'action énergétique).
La combinaison de (13, 16) et l'utilisation de la contrainte (15) conduisent en utilisant le calcul variationnel à l'optimisation des temps de transformation du cycle par rapport à l'énergie mécanique produite :

$$W_{réel} = W_{rev} - (T_{HS} - T_{CS}) \times \left[\frac{\sum_i C_{Ii}}{\zeta_i}\right]$$

Et au lagrangien suivant :

$$L(\zeta_i) = W_{rev} - \Delta T_S \times \sum_i \frac{C_{Ii}}{\zeta_i} - \lambda \times \left[\sum_i \zeta_i - \zeta\right]$$

D'où, l'expression du maximum d'énergie correspondant :

$$MAX_1\, W_{réel} = W_{rev} - \frac{\Delta T_S}{\zeta} \times \left(\sum_i \sqrt{C_{Ii}}\right)^2 \quad (17)$$

et du rendement au maximum de puissance

$$\eta(MAX_1 w) = \eta_C \left(1 - \frac{\left(\sum_i \sqrt{C_{Ii}}\right)^2}{\xi \Delta S}\right)$$

avec $\xi \Delta S$, actions entropique de transfert de cycle

### 2.4 Cas où l'action entropique est une dimension physique finie

Nous avons montré au paragraphe 2.3 que le maximum d'énergie mécanique est associé à une distribution optimale des durées de chaque transformation du cycle, dès lors que la durée totale (ou période de cycle) est fixée (contrainte de dimension en temps fini).

Ce premier maximum en énergie (17) apparaît comme une fonction des actions entropiques de chaque transformation thermodynamique. En imposant que les actions entropiques sont également une dimension physique finie telle que :

$$\sum_i C_{Ii} = C_{IT}$$

L'optimisation par le calcul variationnel conduit alors à l'équipartition des actions entropiques de chaque transformation :

$$C^*_{Ii} = \frac{C_{IT}}{4}$$

$C_{IT}$, actions entropique totale distribuée de façon optimale sur le cycle. Il y correspond le deuxième maximum d'énergie, obtenu de façon séquentielle :

$$MAX_2\, W_{réel} = W_{rev} - 4\frac{\Delta T_S}{\zeta} C_{IT} \quad (18)$$

La conclusion du calcul d'optimisation précédent est que les mécanismes différents de production d'entropie (au moins 2 : irréversibilité d'isothermes et d'adiabatiques) conduisent à une distribution optimale des durées de transformation irréversibles en temps fini, même pour une optimisation en énergie.

De plus, l'introduction du concept d'action entropique, nous a permis de montrer, qu'en présence d'une contrainte de finitude des actions entropiques, l'optimisation en énergie se produit à l'équipartition des actions entropiques des 4 transformations du cycle. Ce résultat est à rapprocher d'autres équipartitions découvertes antérieurement dont la référence [16].

Nous allons maintenant considérer la puissance du moteur thermomécanique.

### 3. **Optimisation en puissance du moteur de CARNOT endo-irréversible**

Il y a lieu ici de noter que, sans les distinguer, la littérature fait apparaître 2 approches formellement différentes. Ainsi les premiers travaux [6,8] portaient sur des systèmes en régime dynamique stationnaire ou permanent au sens du mécanicien des fluides, donc dans lesquels le temps n'intervient pas explicitement. Ainsi beaucoup de publications, se développant dans l'hypothèse d'un régime dynamique stationnaire, se réclament à tort de la Thermodynamique en Temps fini (TTF).

La deuxième approche correspond à celle de la section 2 et se concentre sur un cycle de période $\zeta$, approche similaire à celle de CURZON et AHLBORN [4], mais complétée comme il va être illustré ci-après.

3.1 <u>Synthèse de l'optimisation en énergie</u>

Nous rappelons ici le résultat essentiel obtenu en section 2 :
*En Thermodynamique Irréversible le $MAX\ W$ de l'énergie mécanique est obtenu, au minimum de production d'entropie $\Delta S_I$. Ainsi si $T_{CS} = T_0$, la température ambiante, nous retrouvons le théorème de GOUY-STODOLA [14].*

Par contre cette optimisation en énergie ne concerne qu'un cycle, et non pas le comportement d'un système en régime transitoire. Ceci fera l'objet de futur travaux, déjà entamés [15].

3.2 <u>Optimisation en puissance</u>

La synthèse précédente est prise comme base de départ à l'optimisation en puissance moyenne sur un cycle $\bar{W}$ (et non la puissance instantanée). Cette puissance moyenne est une fonction de $\zeta$, la période du cycle, obtenue à partir de (18) sous la forme :

$$\bar{W} = \frac{W_{rev}}{\zeta} - \frac{4\Delta T_S C_{IT}}{\zeta^2} \qquad (19)$$

Il est facile de voir que cette puissance s'annule pour $\zeta$ tendant vers l'infini (Thermodynamique de l'équilibre.), mais aussi pour $\zeta_{lim}$, période limite telle que :

$$\zeta_{lim} = \frac{4\Delta T_S C_{IT}}{W_{rev}} = \frac{4C_{IT}}{\Delta S_{rev}} = \frac{4C_{IT}}{\Delta S_s}$$

Entre ces 2 valeurs, existe une période $\zeta^*$ (ou une fréquence $x^*$) conduisant au maximum de puissance :

$$\zeta^* = \frac{8\Delta T_S C_{IT}}{W_{rev}} = \frac{8C_{IT}}{\Delta S_s} \qquad (20)$$

Cette période est proportionnelle au coefficient d'action entropique totale $C_{IT}$ et inversement proportionnelle à l'entropie de transfert de chaleur disponible à la source chaude dans les conditions de réversibilité. D'où le maximum en puissance moyenne :

$$MAX\,\bar{W} = \frac{1}{\zeta^*}\left[W_{rev} - \frac{4\Delta T_S C_{IT}}{8 C_{IT}}\Delta S_s\right]$$

$$MAX\,\bar{W} = \frac{\Delta S_s}{8\,C_{IT}}\left[\Delta T_S \Delta S_s - \frac{\Delta T_S \Delta S_s}{2}\right] = \frac{\Delta T_S \Delta S_s^2}{16\,C_{IT}} \qquad (21)$$

Le rendement associé à ce $MAX\,\bar{W}$ vaut :

$$\eta_I\left(MAX\,\bar{W}\right) = \frac{\eta_C}{16} \times \frac{\Delta S_s}{C_{IT}} \times \frac{8\,C_{IT}}{\Delta S_s} = \frac{\eta_C}{2} \qquad (22)$$

On confirme donc, un résultat qui apparaît de façon récurrente dans certains travaux. Le rendement à maximum de puissance moyenne diffère de celui présenté par CURZON-AHLBORN. Il est, selon nous, plus général puisque résultant d'une triple optimisation séquentielle du moteur de CARNOT endo-irréversible (optimisation en énergie par rapport aux $\zeta_i$ ; optimisation par rapport aux coefficients d'action de production d'entropie des transformations du cycle $C_{Ii}$ ; optimisation en puissance moyenne par rapport à $\zeta$ période du cycle).

### 4. <u>Discussion – synthèse des résultats</u>

La présente note, contrairement à la majorité des travaux existants, considère un modèle de moteur thermomécanique de Carnot exo-réversible et endo-irréversible, alors que l'endo-réversibilité est plutôt de mise. Si de plus, on considère les pertes thermiques, on a montré que le rendement au sens du premier principe de la thermodynamique dépend de 2 ratios ayant pour référence l'entropie de transfert de chaleur disponible à la source dans les conditions de réversibilité :

$d_I$ , le degré d'irréversibilité du convertisseur

$d_{LS}$ , le degré de pertes thermiques du système incluant source et puits qui caractérisent le non adiabaticité du système.

Dans le cas du moteur de Carnot endo-irréversible sans pertes thermiques en conservant la même référence entropique, on a montré qu'il existe un premier optimum d'énergie mécanisée par rapport aux durées des transformations isothermes et adiabatiques. Pour ce faire, nous avons introduit le nouveau concept d'action entropique pour chaque transformation. Nous avons obtenu une expression du maximum d'énergie fonction de la durée du cycle et des actions de production d'entropie de chaque transformation.

Une deuxième optimisation de l'énergie mécanisée qui suppose les actions de production d'entropie finies et bornées par une action de production d'entropie totale $C_{IT}$ , conduit à une distribution optimale avec équipartition des actions de production d'entropie de chaque transformation.

L'optimisation en puissance moyenne du moteur de Carnot endo-irréversible a été alors conduite par rapport à la durée du cycle. L'optimum de cette puissance est obtenu pour une période qui vaut :

$$\xi^* = \frac{8\,C_{IT}}{\Delta S_S}$$

La période est donc liée au rapport de l'action de production total d'entropie du cycle divisée par l'entropie de transfert de chaleur disponible dans le cas réversible à la source de chaleur. Le rendement associé au maximum de puissance moyenne vaut alors la moitié du rendement de Carnot, confirmant de façon générale un résultat qui apparaît dans certains travaux. Le résultat diffère fondamentalement de celui obtenu par CURZON et AHLBORN dans le cas endoréversible.

Des prolongements à la présente proposition sont en cours. Elles suivent des communications faites en 2019 à deux colloques internationaux (JETC19, ECOS19).

*Dédicace : cet article a été rédigé en français en hommage à S. Carnot , un des pères de la thermodynamique, et sans doute le scientifique français le plus cité dans le monde grâce à son opuscule [1], œuvre unique de sa courte vie.*